\documentclass[aps,prb,twocolumn,showpacs,superscriptaddress]{revtex4}
\usepackage{graphicx}
\usepackage{bm}

\begin{document}
\newcommand{\beq}{\begin{equation}}
\newcommand{\eeq}{\end{equation}}

\title{Topological crossovers near a quantum critical point}
\author{V.~A.~Khodel}
\affiliation{Russian Research Centre Kurchatov
Institute, Moscow, 123182, Russia}
\affiliation{McDonnell Center for the Space Sciences \&
Department of Physics, Washington University,
St.~Louis, MO 63130, USA}
\author{J.~W.~Clark}
\affiliation{McDonnell Center for the Space Sciences \& Department
of Physics, Washington University, St.~Louis, MO 63130, USA}
\author{M.~V.~Zverev}
\affiliation{Russian Research Centre Kurchatov
Institute, Moscow, 123182, Russia}
\affiliation{Moscow Institute of Physics and Technology, Moscow, 123098, Russia}
\date{\today}
\begin{abstract}

We study the temperature evolution of the single-particle spectrum $\epsilon(p)$
and quasiparticle momentum distribution $n(p)$ of homogeneous strongly correlated
Fermi systems beyond a point where the necessary condition for stability of
the Landau state is violated, and the Fermi surface becomes multi-connected by
virtue of a topological crossover.  Attention is focused on the different
non-Fermi-liquid temperature regimes experienced by a phase exhibiting a
single additional hole pocket compared with the conventional Landau state.
A critical experiment is proposed to elucidate the origin of NFL behavior
in dense films of liquid $^3$He.
\end{abstract}

\pacs{
71.10.Hf, 
71.10.Ay  
67.30.E- 
67.30.hr 
} \maketitle

The study of non-Fermi-liquid (NFL) behavior of strongly correlated
Fermi systems in the regime of a quantum critical point (QCP) is
currently one of the most active and challenging areas of
condensed matter physics. \cite{loh,steglich}  As a rule, such
behavior is attributed to second--order phase transitions, and the
QCP is identified with the end point of a corresponding line of
transition temperatures, denoted by $T_N(H)$ in the prototype in
which an external magnetic field $H$ is the control parameter.  In
this case, NFL behavior is triggered by critical antiferro- or
ferromagnetic fluctuations, which lead to violation of
respective Pomeranchuk stability conditions (PSC).  Ensuing
NFL phenomena are presumably explained either within the
Hertz-Millis  theory\cite{hertz,millis} or, in heavy-fermion
metals, within a Kondo breakdown  model.\cite{loh,steglich,si,pepin2010}

However, the widely promulgated fluctuation
scenario is inconsistent with experimental data on a number of
strongly correlated Fermi systems exhibiting NFL behavior:
\begin{itemize}
\item[(i)]
In dense $^3$He films where the emergent NFL behavior has been
documented, experiment \cite{godfrin1995,godfrin1998,saunders1,saunders2}
has not identified any related second-order phase transition.
\item[(ii)]
In several heavy-fermion metals \cite{bud'ko,stegcol}, concurrent
divergence of the Sommerfeld ratio $\gamma(T)=C(T)/T$ and the
magnetic susceptibility $\chi(T)$ is observed at a point
 that is {\it separated by an intervening NFL
phase} from termination points of any second-order phase transitions.
\item[(iii)]
In many instances of well-pronounced NFL behavior, the order parameters
required to specify associated second--order phase transitions are
still elusive, casting further doubt on the fluctuation scenarios.
\item[(iv)]
In external magnetic fields, thermodynamic properties demonstrate
scaling behavior governed specifically by the ratio $\mu_f H/T$
where $\mu_f$ is the magnetic moment of constituent fermions.
\end{itemize}

These NFL phenomena can be understood when one recognizes
that standard FL theory possesses its own quantum critical point,
in the vicinity of which it fails.  At this point, the {\it necessary}
stability condition (NSC) for the $T=0$ Landau state is
violated,\cite{physrep,khodel2007,prb2008,jetplett2009} as
opposed to violation of some PSC at a conventional QCP.

The NSC states that an arbitrary admissible variation $\delta n( p)$
from the FL quasiparticle momentum distribution $n_F( p)=\theta(p_F-p)$,
while conserving particle number, must produce a positive change of
the ground-state energy $E_0$, i.e.,
\beq
\delta E_0=\int\epsilon(p;n_F( p))
\delta n(p)d\upsilon> 0.
\label{nesc}
\eeq
Here, $\epsilon(p;n_F)$ denotes the spectrum of single-particle
excitations measured from the chemical potential $\mu(T=0)$ and
evaluated for the initial Landau state specified by the
quasiparticle occupancy $n_F(p)$.
The reduction in energy due to breakdown of the NSC, which involves
contributions linear in $\delta n$, is clearly larger than
that due to violation of any PSC, which involves {\it bilinear}
combinations of $\delta n$.  We must conclude that any associated
fluctuation scenario is irrelevant to the different type of QCP
associated with violation of the NSC, which we shall call a
Fermi-liquid QCP.

Violation of the NSC (\ref{nesc}) is {\it unambiguously} linked
to a change of the number of roots of equation
\beq
\epsilon( p,n_F)=0 .
\label{topeq}
\eeq
In standard Fermi liquids, this equation has a
single root at the Fermi momentum $p_F$, and in that case
the signs of $\epsilon(p)$ and $\delta n(p)$ coincide, ensuring
satisfaction of the NSC (\ref{nesc}).  However, consideration
of the full Lifshitz phase diagram anticipates the emergence of
additional roots of Eq.~(\ref{topeq}).  For example, such roots
appear at a critical density $\rho_{\diamond}$ where the function
$\epsilon(p,\rho_{\diamond})$ attains either a maximum, with a bifurcation
point $p_b<p_F$, or a minimum, with $p_b>p_F$, so that
$\epsilon(p\to p_b,\rho_{\diamond})\propto (p-p_b)^2$, (see the upper
two panels of Fig.~1). Thus, vanishing of $\epsilon(p_b,\rho_{\diamond})$
is always accompanied by vanishing of the group velocity
$v(p_b,\rho_{\diamond})=(\partial\epsilon(p,\rho_{\diamond})/\partial p)_{p_b}$.
Beyond the critical density $\rho_{\diamond}$, the NSC fails to hold,
since $\epsilon(p,n_F;\rho)$ and $\delta n(p)$ have opposite signs
close to $p_b$.

As indicated in the lower panel of Fig.~1, the condition (\ref{nesc})
is also violated at a critical density $\rho_{\infty}$ where the
effective mass $M^*(\rho)$ diverges. In this case, standard manipulations
based on
the Landau relation connecting the single-particle spectrum and the quasiparticle momentum distribution, (see Eq.~(\ref{lansp}) below) yield
\beq
{v_F(\rho)\over v^0_F}\equiv {M\over M^*(\rho)}
=1-{1\over 3}F^0_1(\rho)
\label{rel01}
\eeq
where $v^0_F=p_F/M$ and $F^0_1(\rho)=f_1(p_F,p_F;\rho)p_FM/\pi^2$
is the dimensionless first harmonic of the Landau interaction function,
normalized with the density of states $N_0=p_FM/\pi^2$ of the ideal
Fermi gas.  Evidently, $F^0_1(\rho)$ is a smooth function of the
density $\rho$, and $F_1^0(\rho)=3$ at $\rho=\rho_{\infty}$.
Then beyond the critical point, one has $F_1(\rho)>3$, and the
Fermi velocity $v_F(\rho)$ becomes negative.  This behavior conflicts
with the fluctuation scenario for the QCP, in which such a sign
change is impossible.

\begin{figure}[t]
\includegraphics[width=0.55\linewidth,height=0.9\linewidth]{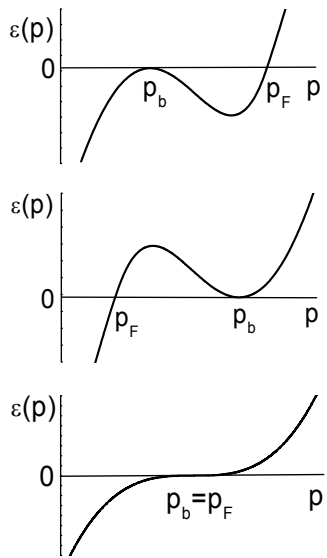}
\caption{
Three scenarios of emergent bifurcation in Eq.~(\ref{topeq}):
$p_b<p_F$ (top panel), $p_b>p_F$ (middle panel), $p_b=p_F$
(bottom panel).
}
\label{fig:bif}
\end{figure}

To summarize, we infer that at any point where the NSC is violated,
the density of states, given by
\beq
N(T)
={1\over T}\int
n( \epsilon)(1-n(\epsilon)){dp\over d\epsilon} d\epsilon,
\label{ds}
\eeq
{\it diverges} at $T\to 0$ due to vanishing of the group velocity $d\epsilon(p)/dp$.
One has\cite{prb2005,prb2008}
\beq
N(T\to 0,\rho_{\infty})\propto T^{-2/3}   , \quad N(T\to 0,\rho_{\diamond})\propto T^{-1/2}  .
\label{n12}
\eeq
The difference in critical indexes is associated with the fact that
$dp/d\epsilon\propto \epsilon^{-2/3}$ at the critical density $\rho_{\infty}$,
whereas $dp/d\epsilon\propto |\epsilon|^{-1/2}$ at the critical density
$\rho_{\diamond}$.

Significantly, the Sommerfeld-Wilson ratio $R_{SW}=\chi(T)/\gamma(T)$
cannot diverge at these points.  Indeed, the density of states $N(T)$
cancels out in the ratio $R_{SW}$, while the Stoner factor entering $\chi(T)$
maintains a finite value, since, as we have seen, the PSC and NSC
cannot fail {\it at the same point}. This conclusion is in agreement
with experimental data\cite{godfrin1995,dolgopolov,stegacta} on
dense films of liquid $^3$He, the two-dimensional electron gas of
MOSFETs, and the majority of heavy-fermion metals.

Since no symmetry is violated at a Fermi liquid QCP, and hence no hidden
order parameters are involved, the transition ensuing from the violation
of the NSC (\ref{nesc}) is {\it topological} in character.\cite{lifshitz,volrev}
Beyond the bifurcation point, Eq.~(\ref{topeq}) usually has two
additional roots $p_1$ and $p_2$ situated near each other (however,
cf.~Refs.~\onlinecite{ks,vol1991,noz}).  It is for variations $\delta n(p)$
involving momenta $p_1<p<p_2$, at which $\delta n(p)$ and $\epsilon(p)$
have opposite signs, that the NSC (\ref{nesc}) breaks down.

The analysis of topological rearrangements triggered by the interaction
between quasiparticles began twenty years ago,\cite{ks} with important
subsequent developments reported in Refs.~\onlinecite{vol1991,noz,
schuck,zb,shagp,zkb,shaghf,prb2005,schofield,ndl,taillefer}.
In this article, we address the Fermi-liquid QCP in homogeneous
matter and focus on the case where the new roots $p_1$ and $p_2$
emerge near the Fermi momentum $p_F$.  The physics of this phenomenon
is captured if we keep the three first terms,
\begin{eqnarray}
\epsilon(x)&=&p_Fx\left(v_F+ {v_1\over 2}x
+{v_2\over 6}x^2\right), \nonumber\\
 v(x)&=&v_F+ v_1x +{v_2\over 2}x^2,
\label{spt}
\end{eqnarray}
in the Taylor expansions of the spectrum $\epsilon(x)$ and its
group velocity $v(x)$, where $x=(p-p_F)/p_F$. To some extent,
this approach is reminiscent of that employed by Landau in his
theory of second--order phase transitions.  In an ideal
Fermi gas, $v_F=v_1=v^0_F=(2M\epsilon^0_F)^{1/2}$.  The case
$v_1=0$, $v_2>0$  was considered in Ref.~\onlinecite{prb2005}.  Here we
assume that $v_1>0$, $v_2>0$, and $v_1/v_2\ll 1$, the situation addressed
in the numerical calculations of Ref.~\onlinecite{prb2008}.

To find the bifurcation momentum $p_b=p_F(1+x_b)$ one must solve
the set of equations $\epsilon(p)=0$ and $v(p)=0$, i.e.
\begin{eqnarray}
v_F+ {v_1\over 2}x_b
+{v_2\over 6}x^2_b&=& 0, \nonumber\\
 v_F+ v_1x_b +{v_2\over 2}x^2_b&=& 0 .
\label{spte}
\end{eqnarray}
This system has the solution $x_b=-3v_1/2v_2$ provided the critical condition
\beq
{8v_2v_F(\rho)\over  3v^2_1}=1
\label{vt}
\eeq
is met. Thus in the case $v_1\neq 0$, the critical Fermi velocity
$v_F$ is still {\it positive}, and therefore the Landau state becomes
unstable {\it before} the system reaches the point at which the
effective mass diverges---as was first discovered and discussed in
Refs.~\onlinecite{zb}.

The prerequisite $x_b\ll1$ for applicability of the expansion (\ref{spt}) is
satisfied provided $v_1/v_2\ll 1$, implying that the critical Fermi velocity
is small: $v_F =3v^2_1/8v_2\ll v^0_F$.  Given this situation, upon accounting
for the dependence of $v_F$ on the temperature $T$ and control parameters such
as the external magnetic field $H$ that do not change the form of Eq.~(\ref{vt}),
one can establish a critical line $T=T_{\diamond}(H)$ separating phases
with {\it different topological structure}.

Evaluation of relevant $T-$ and $H-$dependent corrections to the Fermi
velocity $v_F$ is based on the Landau equation \cite{lan1,lanlif} for the
single-particle spectrum $\epsilon(p)$, which in 3D has the form
\beq
{\partial\epsilon(p)\over\partial p} = {p\over M}
+ {1\over 3}\int\! f_1( p,p_1)\, {\partial n(p_1)\over\partial  p_1}\,
d\upsilon_1, \label{lansp}
\eeq
with $d\upsilon=p^2dp/\pi^2$.  This relation provides a nonlinear integral
equation for self-consistent determination of $\epsilon(p,T,H)$ and the momentum
distribution
\beq
n(p,T,H)=\left[ 1+e^{\epsilon(p,T,H)/T}\right]^{-1},
\label{dist}
\eeq
with the Landau interaction function $f({\bf p},{\bf p}_1)$ (hence its first
harmonic $f_1$) treated as phenomenological input.

\begin{figure}[t]
\includegraphics[width=0.7\linewidth,height=1.2\linewidth]{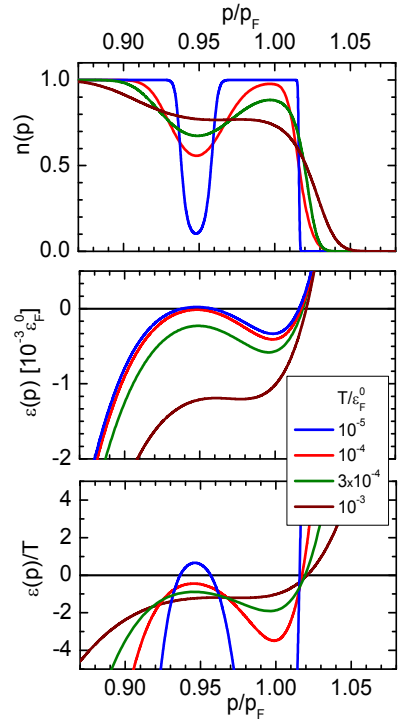}
\caption{
Occupation numbers $n(p)$ (top panel), single-particle spectrum
$\epsilon(p)$ in units of $10^{-3} \epsilon^0_F$ (middle panel), and ratio
$\epsilon(p)/T$ (bottom panel) evaluated for the model (\ref{model_0pf})
with $\kappa=0.07$ and $g_s=0.45$,
at four color-coded temperatures (in units of $\epsilon^0_F$) below $T_*=3
\times 10^{-3}\epsilon^0_F$.
}
\label{fig:top_bub}
\end{figure}

Our goal is to evaluate the $T-$ and $H-$dependence of the key quantity
$v_F(\rho,T,H)$.  In the simplest case $H=0$, the overwhelming $T-$dependent
contributions to $v_F$ come from integration over the vicinity of the
bifurcation momentum $p_b$.
Evaluation is performed along the same lines as in Ref.~\onlinecite{prb2005},
i.e., by expanding the interaction function in a Taylor series,
although here we have to retain a correction to the FL formula (\ref{rel01})
linear in $p-p_b$.
As a result, we arrive at
\beq
v_F(T \to T_{\diamond}) -v_F(\rho)
\propto \int(s-p_b){\partial n(s,T)\over\partial  s} ds,
\label{lanb}
\eeq
where $v_F(\rho)$ is given by Eq.~(\ref{rel01}).

The integral $I_T$ on the right side of Eq.~(\ref{lanb}) is evaluated with
the aid of relations $\epsilon(p\to p_b)\propto (p-p_b)^2$ and
$d\epsilon(p\to p_b)/dp\propto \sqrt{\epsilon(p)}$ stemming from
Eq.~(\ref{spt}).  Upon standard changes of integration variables
$p\to \epsilon\to Tz$, we find
\beq
I_T\propto T^{1/2}\int z^{1/2}n(z)\left(1-n(z)\right) dz\propto \sqrt{T/\epsilon^0_F}
\eeq
at $T\to T_{\diamond}$.
Together with Eq.~(\ref{vt}), this result leads to a tiny value of the critical
temperature
\beq
T_{\diamond}\propto \epsilon^0_F \left({v^2_1\over v_2v^0_F}\right)^2 D^2,
\label{td}
\eeq
where
\beq
D=1-8{v_2v_F^0\over 3v^2_1} \left(1-{1\over 3}F^0_1(\rho)\right) \geq 0
\label{dc}
\eeq
is a criticality parameter.
Temperature  $T_{\diamond}$ vanishes at the point where $D=0$.

Consider now the imposition of a magnetic field $H$ on the system.
The impact of the field becomes well
pronounced when $\mu_f H>T$, and there emerge two subsystems
having spin projections $\pm 1/2$, implying in turn
a decomposition $N(T)=N_+(T)+N_-(T)$ of the density of states.
The corresponding formulas are cumbersome and will be analyzed elsewhere.
Here we focus on the case $T=0$ and estimate
an upper tuning magnetic field
$H_{\diamond}$ such that a bifurcation emerges in the spectrum
$\epsilon_+(p)$, while the down-spin spectrum
$\epsilon_-(p)=\epsilon_+(p)-2\mu_f H$
admits merely the conventional root $p_F^-$.  As before, a leading correction
$I_H$ to $v_F$ comes from integration over the vicinity of the momentum $p_b$,
with the subsystem whose spectrum goes to
$\epsilon_+(p)\propto (p-p^+_b)^2
=Tz+\mu_f H_{\diamond}$
at $H\to H_{\diamond}$ making the dominant contribution, to yield
\beq
I_H\propto \int \left(Tz+\mu_f H_{\diamond}\right)^{1/2}n(z)\left(1-n(z)\right) dz\propto \sqrt{\mu_f H_{\diamond}/ \epsilon^0_F}
\eeq
and
\beq
\mu_f H_{\diamond}\propto \epsilon^0_F\left({v^2_1\over v_2v^0_F}\right)^2 D^2 .
\label{hd}
\eeq
Comparing Eqs.~(\ref{td}) and (\ref{hd}) we see that
$\mu_f H_{\diamond} \sim T_{\diamond}$.  This result is inherent to a scenario
in which single-particle degrees of freedom play the dominant role and is
consistent with available experimental data on heavy-fermion
metals.\cite{stegacta,krellner}

Let us now briefly analyze the situation at $T=H=0$ on the ordered side of the
topological rearrangement
assuming, as before, the criticality parameter $D$ to be positive.
In the case $v_1>0$, addressed first in Refs.~\onlinecite{zb} and later in
Ref.~\onlinecite{prb2008}, the bifurcation momentum $p_b$ resides
inside the Fermi volume. The rearranged $T=0$ quasiparticle momentum
distribution $n(p)$ is given by $n(p)=1$ for $p<p_1$ and $p_2<p<p_F$,
and zero otherwise, with $p_F$ shifted outward to conserve quasiparticle
number.  Thus, the Fermi surface gains an additional
hole pocket. In the 1960's, such a small hole pocket was called a Lifshitz
bubble (LB) in Landau-school folklore.  In this case, two additional roots
of Eq.~(\ref{topeq}) appear, with
\beq
x_{1,2}= -{3v_1\over 2v_2}\left(1\pm  \sqrt{1-8{v_2v_F(\rho;p_1,p_2)\over 3v^2_1}} \right).
\label{rob}
\eeq
We note that $v_F(\rho,p_1,p_2)$ differs from the parameter $v_F(\rho)$ introduced
previously, since it is evaluated for the phase in which the Fermi surface has three sheets.
Accounting for the displacement of $p_F$ due to emergence of the LB, one obtains
$v_F(\rho,p_1,p_2)-v_F(\rho)\propto (p_1-p_2)^2$, leading to
\beq
p_2-p_1\propto {v_1\over v_2}\sqrt{D}.
\eeq
As a result, we find
\beq
v_{LB}\propto{v_1^2\over v_2}\sqrt {D}< v_F
\eeq
for the LB Fermi velocity $v_{LB}=v(x_1)$ from the second of Eqs.~(\ref{spt}),
thereby demonstrating that the LB contribution to the density of states $N(0)$ prevails.

At temperatures beyond $T>T_{\diamond}$, the LB contribution to thermodynamic
properties disappears.  Were this to occur instantaneously, the specific
heat $C(T)$ would undergo a jump, as if one were dealing with a second--order
phase transition.  As a matter of fact, the rearrangement occurs rapidly but
not momentarily. Thus one deals with a {\it topological crossover} (TC), and
Eqs.~(\ref{spte}) serve to establish a TC line $T_{\diamond}(H)$ that resembles
a line $T_N(H)$ of second--order phase transitions.

The TC width is found from the condition $T_< <T_{\diamond}<T_>$,
with the boundaries $T_<$ and $T_>$ being determined by the relations
\beq
\epsilon(p_b,T_<)=T_{\diamond}, \quad
\epsilon(p_b,T_>)=-T_{\diamond}  \  .
\label{bo}
\eeq
Since $v(p,T)\simeq v_{\diamond}(p)\sqrt{T/\epsilon^0_F}$
in the LB region near $T_{\diamond}$, the similar formula
$\epsilon(p,T)=(\epsilon_{\diamond}(p)-\epsilon_{\diamond}(p_F))\sqrt{T}$
is obtained for the spectrum $\epsilon(p,T)$ after a simple momentum
integration. Straightforward manipulations employing the definition
$\epsilon(p_b,T_{\diamond})=0$ then lead to
\beq
{T_>-T_{\diamond}\over T_{\diamond}}\simeq
{T_{\diamond}-T_<\over T_{\diamond}}
\propto \sqrt{T_{\diamond}/\epsilon^0_F}.
\eeq
Accordingly, the reduced temperature width of the critical region turns
out to be small, implying that the TC does indeed imitate a second--order phase
transition.

A conventional FL regime having $T$-independent quantities $\chi(0)\propto
\gamma(0)\propto N(0)\propto 1/v_{LB}(0)\simeq 1/[v_F(0)\sqrt{D}]$
is seen to persist until $T$ reaches $T_< < T_{\diamond}$, where
the LB occupation numbers begin to experience substantial
change as the temperature continues to increase.  Both the density of states $N(T)$
and the spin susceptibility $\chi(T)$ attain maximum values at $T=T_{\diamond}$, where
$\chi(T_{\diamond})\propto 1/(v(x_b,T_{\diamond}))\propto {1/ \sqrt{T_{\diamond}}}$.
At higher temperatures, the LB contribution to $\chi(T)$ begins to fall,
finally dying out and leaving $\chi(T)\propto 1/v_F(0)$.  Analogous results
are found for the Sommerfeld ratio, given by
\beq
\gamma(T)=
\int {\epsilon(p)\over T}{\partial n(p)\over \partial T}d\upsilon,
\label{fcct}
\eeq
except that $\gamma$ reached its maximum at a different temperature, due to the marked
dependence of the spectrum $\epsilon(p,T)$ on $T$.  The foregoing analysis
therefore leads to the conclusion that in the QCP region,
both the magnetic susceptibility $\chi(T)$ and the Sommerfeld ratio $\gamma(T)$
exhibit {\it asymmetric peaks}, located at different temperatures
$\simeq T_{\diamond}$.  Such behavior, observed in many heavy-fermion metals
situated in a QCP region,\cite{stegacta,krellner,steg2011} remains unexplained
within any conventional scenario for the QCP.

\begin{figure}[t]
\includegraphics[width=0.63\linewidth,height=1.16\linewidth]{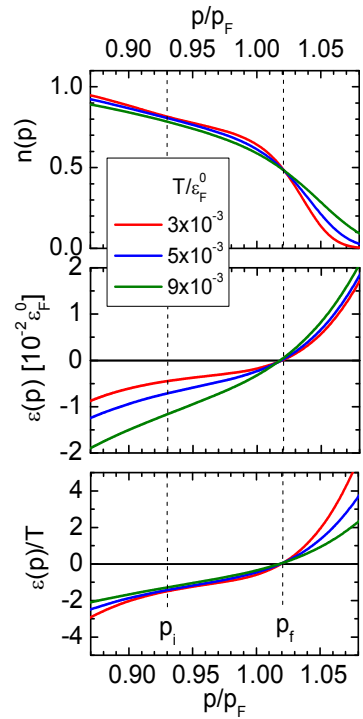}
\caption{
Same as in Fig.~\ref{fig:top_bub} but at $T\geq T^*$.
}
\label{fig:top_fc}
\end{figure}

Further temperature evolution of the spectrum $\epsilon(p,T)$ is associated
with another essential rearrangement\cite{prb2008} of the momentum distribution
$n(p,T)$ that occurs in the region of a critical temperature
$ T_*$.  The distribution $n(p,T)$ becomes a smooth function of momentum
\beq
n(p,T)\simeq n_*(p) ,  \quad  p_i<p<p_f,
\label{dec}
\eeq
in an interval adjacent to the Fermi surface and is
otherwise unity for $p<p_i$ and zero for $p>p_f$.
In this domain, $n(p,T)$ is nearly independent of $T$, while the dispersion of
the single-particle spectrum $\epsilon(p,T)$ becomes proportional to $T$
so as to satisfy Eq.~(\ref{dist}).

\begin{figure}[t]
\includegraphics[width=0.65\linewidth,height=1.05\linewidth]{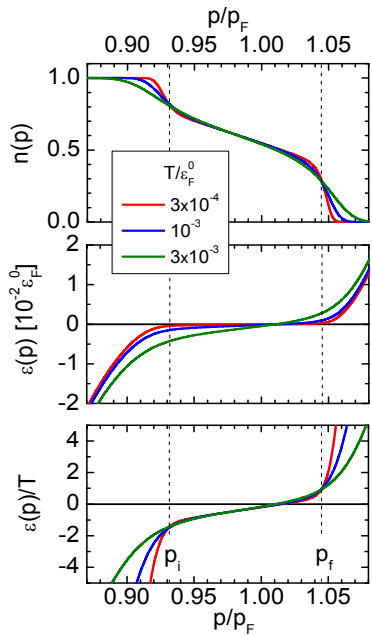}
\caption{
Same as in Fig.~\ref{fig:top_fc} but for the  interaction function (\ref{model_yuka})
with the parameters $\alpha=10$ and $g_y=70$ at three
line-type-coded temperatures.
}
\label{fig:fc_yuka}
\end{figure}

Both these features are inherent to the phenomenon
of fermion condensation, a {\it topological} phase transition discovered
twenty years ago,\cite{ks,vol1991,noz,physrep,shagrev,shagrep}
in which a flat band pinned to the Fermi surface (the so-called
fermion condensate (FC)) is formed.  This phenomenon, alternatively
viewed as a swelling of the Fermi surface, was recently rediscovered
by Lee \cite{lee} while investigating the finite-charge-density
sector of conformal field theory (CFT) within the AdS/CFT
gravity/gauge duality.
The phenomenon of fermion condensation (flat band) may also arise in
topological media for purely topological reasons, 
(see Refs.\onlinecite{vol2011a,vol2011b}).

\begin{figure}[t]
\includegraphics[width=0.84\linewidth,height=0.92\linewidth]{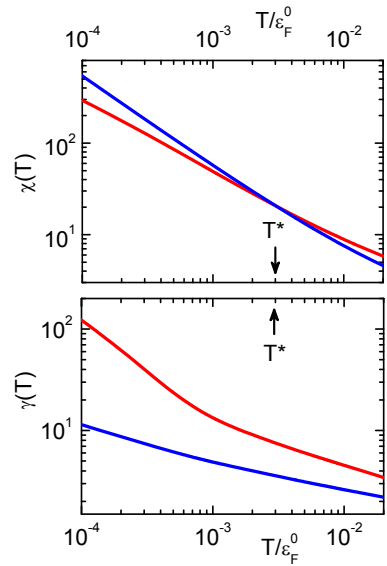}
\caption{
Spin susceptibility $\chi(T)$ (top panel) and Sommerfeld ratio
$\gamma(T)=C(T)/T$ (bottom panel), in Fermi gas units
$p_FM/\pi^2$ and $p_FM/3$, respectively. Red curves show results for the
model (\ref{model_0pf}) and blue curves, for the model (\ref{model_yuka}).
}
\label{fig:thermo}
\end{figure}

Unfortunately, important details of this rearrangement cannot be
established analytically. To clarify the relationship between properties of
the phase having the single LB at $T=0$ and those of a system possessing
a FC at $T=0$, we must resort to numerical treatment of Eq.~(\ref{lansp}).
Figs.~\ref{fig:top_bub} and ~\ref{fig:top_fc} present results from numerical
calculations\cite{prb2008} of the spectra $\epsilon(p)$ and momentum distributions
$n(p)$ for a 3D model system based on the interaction function
\beq
f(q)=g_s{\pi^2p_F\over M}{1\over q^2+\beta^2p^2_F}
\label{model_0pf}
\eeq
with dimensionless parameters $g_s=0.45$ and $\beta=0.07$, values for which
the zero-$T$ phase possesses a single LB.  In this model one has
$T_{\diamond}\simeq 5 \times 10^{-5}\epsilon^0_F$ and $T_*=3\times 10^{-3}\epsilon^0_F$.
The results are to be compared with those in Fig.~\ref{fig:fc_yuka}
obtained for the model interaction function\cite{prb2008}
\beq
f(q)=g_y{\pi^2\over M}{e^{-\alpha q/p_F}\over q},
\label{model_yuka}
\eeq
for which a flat portion in the spectrum $\epsilon(p)$ is already present at $T=0$.
In the interval $T\simeq T_{\diamond}<T_*$, the spectra $\epsilon(p)$ of the
two systems are quite dissimilar.  However, when $T$ reaches values
around $ T_*$, a flat portion of $\epsilon(p)$ develops for the interaction
model (\ref{model_0pf}) as well, the density associated with the
 flat segment being half the FC density $\rho_*$ obtained for the model
(\ref{model_yuka}).  On the other hand,
 outside the range $[p_i,p_f]$,
the momentum distribution $n(p)$ calculated for model (\ref{model_0pf})
shows more pronounced tails than in the case of model (\ref{model_yuka}).

\begin{figure}[t]
\includegraphics[width=0.8\linewidth,height=0.94\linewidth]{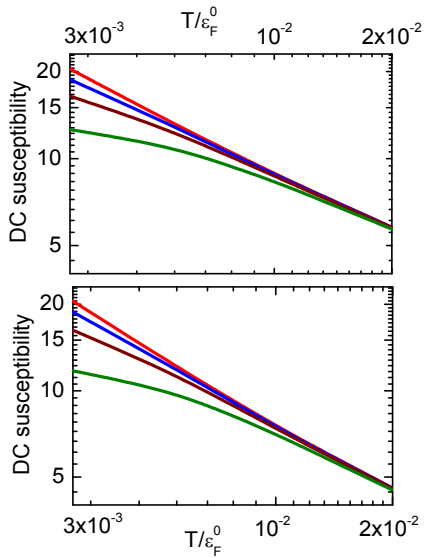}
\caption{
DC susceptibility (in Fermi gas units) as a function of
temperature (in units of $\varepsilon^0_F$) for different external-field
magnitudes. Colors correspond to different values of
$\mu_f H/\varepsilon^0_F$: $10^{-3}$ (red curve), $3 \times 10^{-3}$ (blue
curve), $5 \times 10^{-3}$ (brown curve), $9 \times 10^{-3}$ (green curve).
Top panel: model (\ref{model_0pf}), bottom panel: model
(\ref{model_yuka}).
}
\label{fig:magn}
\end{figure}

The impact of these differences on the magnetic susceptibilities $\chi$
and Sommerfeld ratios $\gamma(T)$ of the two model systems is seen in
Fig.~\ref{fig:thermo}.  As is known,\cite{zk4,yak} the contribution
$\chi_*$ of the FC region to $\chi$ obeys the Curie law:
$\chi_*(T)\propto C_*/ T$, with an effective Curie constant
\beq
C_*\propto\int n_*(p)(1-n_*(p))d\upsilon\propto \rho_*  .
\eeq
Fig.~\ref{fig:thermo} shows that this Curie-like term does prevail in the
susceptibilities calculated for both models.  Such NFL behavior is in agreement
with experimental data on dense films of liquid $^3$He obtained in the QCP
region.\cite{godfrin1995,godfrin1998,saunders2}.

The Sommerfeld ratios $R_{SW}=\chi(T)/\gamma(T)$ evaluated for the two models
differ drastically.  Indeed, in the LB case (i.e., for model (\ref{model_0pf})),
the values of $\chi(T\to 0)$ and $\gamma(T\to 0)$ are large compared with those
of the corresponding ideal Fermi gas, with $R_{SW}$ remaining of order of unity.
On the other hand, in the model (\ref{model_yuka}) that already hosts a FC at
$T=0$, $\chi(T )$ diverges at $T\to 0$ as $T^{-1}$, while the FC contribution
to $\gamma(T)$ evidently vanishes, implying a huge enhancement of $R_{SW}$.
This scenario is in agreement with data\cite{steg2005} on the specific heat
$C(T)$ of the $P$-type heavy-fermion metal YbIr$_2$Si$_2$, for which
the FL term $C(T)\propto T$ exists only at extremely low temperatures
below $T_{\diamond}$, whose value is presumably less than 1 ${\rm K}$.  The
corresponding value of $\gamma(T\to 0)$ is so enhanced that already at
$T\simeq 0.7K$, the entropy value has become surprisingly large:
$S/N\simeq (\ln 2)/2$.  At $T>T_{\diamond}$ a collapse of $C(T)$ occurs,
and the Landau term linear in $T$ {\it completely disappears}.  Thus,
the $P$-type of the compound YbIr$_2$Si$_2$ presents the first example of
a new class of metals, in which flattening of the single-particle spectrum
results in an ordinary NFL shape of $C(T)$ at extremely low $T$.

Next, recall that in conventional Fermi liquids, the value of the susceptibility
$\chi(T,H)$ is proportional to the density of states at finite field $H$, but
is almost independent of $H$.  By contrast, as witnessed in Fig.~\ref{fig:magn},
an external magnetic field suppresses the NFL contribution to $\chi$
in Fermi systems whose spectra $\epsilon(p)$ exhibit a flat region
(and to the same extent independent of which model is chosen).
The larger the magnitude of
the dimensionless parameter $\mu_f H/T$, the more pronounced is the
suppression.  These results, evaluated within the scheme elaborated in
Refs.~\onlinecite{schuck2003,shaghf,prb2005}, elucidate the NFL
behavior of dense liquid--$^3$He films reported in
Refs.~\onlinecite{godfrin1995,godfrin1998,saunders1,saunders2}.
Taking for $H$ a typical value of 1 T and for $\mu_f$ the
magnitude of the $^3$He atom's magnetic moment, the
inequality $\mu_f H>T$ is met at $T<0.5$ mK.  Since temperatures below
$0.2$ mK are currently attainable, we suggest that it is experimentally
feasible to verify or refute the predicted existence of a domain of the
$(T,H)$ phase diagram of 2D liquid $^3$He sensitive to the magnitude of $H$.

We have investigated topological transitions arising in
strongly correlated Fermi systems beyond a quantum critical point of the
Fermi-liquid type, at which the density of states diverges while the
Sommerfeld-Wilson ratio  remains finite. We have
attributed these transitions to violation of the {\it necessary} condition
for stability of the Landau state.
 We have shown that in the QCP density region, the relevant phase
diagram features different topological crossovers, occurring
between states of the same symmetry but with different numbers
of sheets of the Fermi surface.  Importantly, the QCP scenario
based on topological crossovers does not entail any order parameters;
hence it is free from the persistent ambiguity of conventional
fluctuation scenarios associated with hidden order parameters.
Our analysis predicts the existence of a domain of the $(T,H)$
phase diagram of 2D liquid $^3$He that is sensitive to the
magnitude of the magnetic field.
This prediction is subject to experimental test.

The authors are grateful to E.\ Abrahams, H.\ Godfrin and  V.\ Shaginyan
for numerous helpful discussions.  This research was supported by the McDonnell Center for the Space Sciences, by Grant Nos.~2.1.1/4540 and NS-7235.2010.2 from the Russian Ministry of Education and Science, and by Grant No.~09-02-01284 from the Russian Foundation for Basic Research.

 \end{document}